\RequirePackage{fix-cm}
\documentclass[smallcondensed]{svjour3}     % onecolumn (ditto)
\smartqed  % flush right qed marks, e.g. at end of proof
\usepackage{graphicx}
\usepackage{morefloats}
\usepackage{amsmath}
\usepackage{amssymb}
\usepackage[algoruled]{algorithm2e}
\usepackage[hidelinks,breaklinks]{hyperref}
\DeclareMathOperator{\entrywisemax}{MaxEntry}
\SetAlgoSkip{smallskip}
\SetAlgoInsideSkip{smallskip}
\SetAlCapFnt{\small}
\SetAlCapNameFnt{\small}
\journalname{arXiv}
\begin{document}

\title{Randomized algorithms for distributed computation
       of principal component analysis and singular value decomposition
       \thanks{Y. Kluger and H. Li were supported in part
       by United States National Institutes of Health grant 1R01HG008383-01A1.
       Y. Kluger is with the Program in Applied Mathematics,
       the Program in Biological and Biomedical Sciences, the Cancer Center,
       the Center for Medical Informatics, and the Department of Pathology
       in the School of Medicine at Yale University.
}}

\titlerunning{Randomized algorithms for distributed computation of PCA and SVD}

\author{Huamin Li \and Yuval Kluger \and Mark Tygert}


\institute{H. Li \at
              Yale University, Program in Applied Mathematics,
              51 Prospect St., New Haven, CT 06510\\
              \email{huamin.li@yale.edu}           
           \and
           Y. Kluger \at
              Yale University, School of Medicine, Department of Pathology,
              Suite 505L, 300 George St., New Haven, CT 06520\\
              \email{yuval.kluger@yale.edu}
           \and
           M. Tygert \at
              Facebook Artificial Intelligence Research,
              1 Facebook Way, Menlo Park, CA 94025\\
              \email{tygert@fb.com}
}

\date{\vspace{-.75in}}%%%%%%%%Received: date / Accepted: date}

\maketitle

\begin{abstract}
Randomized algorithms provide solutions to two ubiquitous problems:
(1) the distributed calculation of a principal component analysis
or singular value decomposition of a highly rectangular matrix, and
(2) the distributed calculation of a low-rank approximation (in the form
of a singular value decomposition) to an arbitrary matrix.
Carefully honed algorithms yield results that are uniformly superior
to those of the stock, deterministic implementations in Spark
(the popular platform for distributed computation); in particular,
whereas the stock software will without warning return
left singular vectors that are far from numerically orthonormal,
a significantly burnished randomized implementation generates
left singular vectors that are numerically orthonormal
to nearly the machine precision.
\end{abstract}

\section{Introduction}

Singular value decomposition is the ``Swiss Army Knife'' and ``Rolls Royce''
of matrix decompositions due to its ubiquitous, unsurpassed utility,
as observed by~\cite{golub-mahoney-drineas-lim}, quoting Dianne O'Leary.
Indeed, singular value decomposition provides highly accurate solutions
to most problems in numerical linear algebra, and is particularly important
for low-rank approximation and principal component analysis,
which is among the most important methods in statistics, data analysis
and analytics, and machine learning, as discussed by~\cite{jolliffe}
and many others. The singular value decomposition of a matrix $A$ consists
of matrices $U$, $\Sigma$, and $V$ such that
\begin{equation}
A = U \Sigma V^*,
\end{equation}
where the columns of $U$ are orthonormal, as are the columns of $V$,
the adjoint (conjugate transpose) of $V$ is $V^*$,
and $\Sigma$ is square and diagonal and its entries are nonnegative
(this is called the ``economic,'' ``reduced,'' or ``thin''
singular value decomposition by~\cite{golub-van_loan}).
Accurate, efficient numerical calculation of the singular value decomposition
is essential for many applications, including in the now common setting
that the data being processed is distributed across multiple computers.

In fact, distributed calculations on clusters of computers
are spawning whole industries devoted
to software platforms such as Hadoop and Spark,
as discussed by~\cite{linden-krensky-hare-idoine-sicular-vashisth},
\cite{gittens-et_al}, and many others;
Spark is rapidly becoming a dominant software platform,
enabling distributed machine learning via its MLlib library.
The current version of Spark's MLlib includes rudimentary routines
for calculating the singular value decomposition;
we aim to make improvements based on combining and honing both well-known
and lesser-known numerical methods and tricks.
The concluding sentence of Section~\ref{conclusion} below lists
several often-overlooked details that turn out to be especially important
(and should make sense courtesy of Sections~\ref{tssvd} and~\ref{low-rank}).

Below, we consider two problems that are of particular interest
for distributed computation due to their computational tractability:
\{1\} calculating the ``economic'' (that is, the ``reduced'' or ``thin'')
singular value decomposition of a tall and skinny matrix
(a matrix for which a full row can fit on a single machine), and
\{2\} calculating a low-rank approximation to an arbitrary matrix such that
the spectral norm of the difference between the approximation
and the matrix being approximated is nearly as small as possible.
These problems are relatively tractable since every matrix
in the resulting decompositions has at least one dimension that is small enough
to ensure that a full row can fit in memory on a single machine.
The solutions below to the second problem \{2\} leverage
the solutions to the first problem \{1\}, returning all results
in the form of a singular value decomposition $U \Sigma V^*$,
where the columns of $U$ are orthonormal, as are the columns of $V$,
and $\Sigma$ is diagonal and its entries are nonnegative.

The rest of the present paper has the following structure:
Section~\ref{tssvd} addresses problem \{1\},
introducing Algorithms~\ref{tsSVD}--\ref{GramSVD2}.
Section~\ref{low-rank} addresses problem \{2\},
introducing Algorithms~\ref{subspits}--\ref{combob},
which leverage the Algorithms~\ref{tsSVD}--\ref{GramSVD2} discussed
in Section~\ref{tssvd}. Section~\ref{conclusion} summarizes our outlook.
Appendices~\ref{only18} and~\ref{devil} corroborate the results
presented earlier in the paper,
regardless of the number of executors in our cluster of computers.
Appendix~\ref{gentimes} displays the times required to synthesize the matrices
in our tests (just for reference, for comparative purposes).
None of the appendices is integral to the main points of the present paper,
and may be omitted.

\begin{remark}
\label{working}
Throughout the present paper, the ``working precision'' refers
to the machine precision adjusted to account for roundoff error.
We simply set the working precision a priori,
based on how much roundoff we might tolerate.
For instance, the working precision could be
$10^{-11}$ for double-precision floating-point arithmetic with matrices
of the sizes considered below (whereas the machine precision would be
$2.2 \times 10^{-16}$). When interpreting the tables, please note
the italicized text in Table~\ref{heads}
for the heading $\| A - U \Sigma V^* \|_2$.
\end{remark}

\begin{remark}
\label{goal}
The main purpose of our implementation is to add efficient
principal component analysis and singular value decomposition to Spark,
not to compute them as efficiently as could be possible
(in fact, bypassing Spark may enhance efficiency). On the whole,
the aspects of Spark unrelated to sophisticated mathematical algorithms
tend to be more important than the parts dependent on such algorithms,
even though the algorithmic aspects are the subject of the present paper.
As much as we would like to think that our own contributions are
the most important, we do realize that database management, stream processing,
fault tolerance and recovery, ease of deployment and administration,
coupling with other systems, etc.\ are typically far more important.
Spark is becoming a dominant platform for machine learning at scale,
so the purpose of our implementation is to enable big data analytics for Spark,
whether or not Spark is the ideal platform
for principal component analysis or singular value decomposition.
In particular, Spark is likely to distribute data over clusters as appropriate
for tasks other than principal component analysis
and singular value decomposition, and our implementation
must deal with the data as distributed however Spark sees fit.
This remark is especially important in light of the findings
of~\cite{gittens-et_al}.
\end{remark}

\begin{remark}
The Spark implementation is available at http://github.com/hl475/svd
(and we hope that the main branch of Spark will pull in these changes soon).
For expository purposes, we also provide a serial implementation in Python 3,
at http://tygert.com/valid.tar.gz (the algorithms of the present paper
are meant for parallel computation, but we opt to provide the Python 3 codes
in addition to the implementation for Spark, as the Python is far easier
to read and run).
\end{remark}

\begin{remark}
Many of the algorithms discussed below are randomized. Exhaustive prior work,
including that of~\cite{ballard-demmel-dumitriu}, \cite{demmel-dumitriu-holtz},
\cite{demmel-grigori-hoemmen-langou}, \cite{halko-martinsson-tropp},
\cite{li-linderman-szlam-stanton-kluger-tygert}, 
\cite{yamazaki-tomov-dongarra1}, and~\cite{yamazaki-tomov-dongarra2},
demonstrates that the randomized methods are at least as accurate,
reliable, and efficient as the more classical deterministic algorithms.
The probability of obtaining results departing significantly
from those observed is well known to be negligible, both empirically
and theoretically (for rigorous proofs, see the work mentioned
in the previous sentence).
\end{remark}

\section{Thin singular value decomposition of tall and skinny matrices}
\label{tssvd}

There are many ways to calculate the singular value decompositions
of tall and skinny matrices (matrices for which a full row can fit
on a single machine). This section compares different methods.

One method is perfectly numerically stable --- producing results
accurate to nearly the machine precision --- but requires
merging intermediate results through multiple levels of a dependency tree
(such a merge is known as a ``reduction''); this is the randomized method
of~\cite{ballard-demmel-dumitriu} and Section~5 of~\cite{demmel-dumitriu-holtz}
(which draws on the techniques of~\cite{demmel-grigori-hoemmen-langou}
and others). The randomization there eliminates the need for pivoting.
The pseudocode in Algorithms~\ref{tsSVD} and~\ref{tsSVD2} summarizes
this randomized method, accounting for several considerations discussed
shortly. Another method, based on computing the Gram matrix $A^* A$
of the matrix $A$ being processed, loses half the digits in some cases,
but can leverage extremely efficient accumulation/aggregation strategies
with minimal blocking dependencies and synchronization requirements;
this is the method of~\cite{stathopoulos-wu},
\cite{fukaya-nakatsukasa-yanagisawa-yamamoto}, \cite{yamazaki-tomov-dongarra1},
and~\cite{yamazaki-tomov-dongarra2}.
The pseudocode in Algorithms~\ref{GramSVD} and~\ref{GramSVD2}
summarizes this method, accounting for the following considerations.
Both methods produce highest accuracy when running their orthonormalization
of singular vectors twice in succession, though running twice is superfluous
during all but the last of the subspace iterations of the randomized algorithms
for low-rank approximation discussed in the following section.
Running the orthonormalization twice ensures that the resulting
singular vectors are numerically orthonormal, though running twice
has little effect on the accuracy of their linear span (and little effect
on the spectral norm of the difference between the calculated decomposition
and the matrix $A$ being processed).

A third method, based on ``tournament pivoting,''
is similar to the first method discussed above,
being reasonably numerically stable while requiring the merging
of intermediate results through multiple levels of a dependency/reduction tree;
this is the method of~\cite{demmel-grigori-gu-xiang} and others.
This third method is deterministic (unlike the first method),
but is otherwise more complicated, less efficient, less accurate,
and weaker theoretically with regard to revealing the ranks
of the rank-deficient matrices commonly encountered
when using the singular value decomposition for dimension reduction;
similar remarks apply to a fourth method, that of~\cite{benson-gleich-demmel}.
In the sequel, we consider only the first two methods discussed above.
These two methods also happen to be the easiest to implement
(which can be critical for future deployment and maintenance).

\begin{remark}
\label{randomness}
For convenience, in our implementation of the first method mentioned above
(the randomized method) we replaced the usual random Gaussian matrix
with a product $D F S \tilde{D} F \tilde{S}$, where $D$ and $\tilde{D}$
are diagonal matrices whose diagonal entries are independent
and identically distributed random numbers drawn uniformly
from the unit circle in the complex plane,
$F$ is the discrete Fourier transform,
and $S$ and $\tilde{S}$ are independent uniformly random permutations
(calculated via the Fisher-Yates-Durstenfeld-Knuth shuffle
of~\cite{durstenfeld}).
To process vectors of real numbers (rather than complex numbers),
we partitioned the vectors into pairs of real numbers and viewed each pair
as consisting of the real and imaginary parts of a complex number.
We found empirically that chaining two products $D F S$
into $D F S \tilde{D} F \tilde{S}$ was sufficient;
chaining a few (specifically, logarithmic in the number of columns
of the matrix whose singular value decomposition is being computed)
is rigorously known to be sufficient, as proven by~\cite{ailon-rauhut}.
Chaining several is affordable computationally but seems like overkill.
\end{remark}

\begin{remark}
\label{normalizing}
When testing the second method mentioned above (the method based
on the Gram matrix) we found that explicitly normalizing
the left singular vectors improved accuracy significantly.
Explicitly normalizing does require computing the Euclidean norms
of the columns of the matrix of left singular vectors, but this costs
substantially less than computing the Gram matrix in the first place.
\end{remark}

\begin{remark}
For TSQR of~\cite{demmel-grigori-hoemmen-langou} used
in Algorithms~\ref{tsSVD} and~\ref{tsSVD2}, we modified
Spark's stock implementation of TSQR to be numerically stable
for any (possibly rank-deficient) input matrix.
\end{remark}

The remainder of the present section gives empirical results
on the first two methods discussed above
and on the existing thin singular value decomposition
of tall and skinny matrices in Spark's MLlib, which is based on Gram matrices,
similar to the second method mentioned above (though the existing Spark routine
lacks the refinement of Remark~\ref{normalizing}).
We demonstrate the performance of the first two methods both when running
their orthonormalization twice in succession and when running only once
(with the resulting loss of accuracy and gain in speed).
Our primary implementations of the algorithms in the present section process
an IndexedRowMatrix from Spark's MLlib.

Our software includes examples of matrices with many different distributions
of singular values and singular vectors. For clarity of the presentation,
the results we present below pertain to the following class of matrices ---
matrices for which the various algorithms produced accuracies near the worst
that we encountered in our experiments:

\begin{equation}
\label{originalmat}
A = U \Sigma V^*,
\end{equation}
where $U$ and $V$ are $m \times m$ and $n \times n$ discrete cosine transforms,
respectively, and $\Sigma$ is the $m \times n$ matrix whose entries
are all zeros aside from the diagonal entries
\begin{equation}
\label{testS}
\Sigma_{j,j} = \exp\left( \frac{j-1}{n-1} \cdot \ln\left(10^{-20}\right) \right)
\end{equation}
for $j = 1$, $2$, \dots, $n$.
Note that, like $A$ from~(\ref{originalmat}) and~(\ref{testS}),
matrices arising from real data are often numerically rank-deficient;
indeed, real data sets are often messy, with duplicate or nearly duplicate
columns and rows, symmetries or near symmetries that limit the numerical rank,
etc. Singular value decomposition and principal component analysis
are very helpful for untangling the mess in real data, and certainly need
to function reliably in such circumstances, circumstances such that
the matrix being processed may be highly ill-conditioned.

The headings of the tables have the meanings detailed in Table~\ref{heads}.
Our Spark environment is detailed in Table~\ref{sparkset}.
We used many --- 20 --- iterations of the power method in order to ascertain
the spectral-norm errors reported in the tables.
The timings in the tables do not include the time spent checking the accuracy
(we used so many power iterations just to be extra careful in providing
highly accurate error estimates, in order to facilitate fully trustworthy
comparisons of the different algorithms).

Tables~\ref{tab1234big}, \ref{tab1234med}, and~\ref{tab1234tiny} report
timings and errors for several experiments.
The reconstruction errors $\|A-U \Sigma V^*\|_2$
for Algorithms~\ref{tsSVD} and~\ref{tsSVD2} (which are similar)
are clearly superior to all those
for Algorithms~\ref{GramSVD} and~\ref{GramSVD2},
which makes sense since the latter algorithms use the Gram matrix
and can therefore lose half their digits.
For the left singular vectors, the errors $\entrywisemax(|U^* U - I|)$
for Algorithms~\ref{tsSVD2} and~\ref{GramSVD2} (which are similarly good)
are clearly superior to all those
for Algorithms~\ref{tsSVD} and~\ref{GramSVD}, which makes sense since
the latter algorithms orthonormalize the singular vectors only once.
For the right singular vectors, the error $\entrywisemax(|V^* V - I|)$
is near the machine precision ($2.2 \times 10^{-16}$) for all algorithms.
All together, then, Algorithm~\ref{tsSVD2} is the most accurate of all,
with all its errors approaching the machine precision adjusted for roundoff.
However, on our cluster with our version of Spark,
Algorithm~\ref{GramSVD2} is somewhat faster than Algorithm~\ref{tsSVD2};
the reconstruction error $\|A-U \Sigma V^*\|_2$ is somewhat worse
for Algorithm~\ref{GramSVD2} than for Algorithm~\ref{tsSVD2},
but may be acceptable in many circumstances.

As expected, the timings in Tables~\ref{tab1234big}, \ref{tab1234med},
and~\ref{tab1234tiny} are roughly proportional to the numbers of rows
in the matrices (the number of columns is fixed throughout these tables),
and the errors adhere to the working precision
mentioned in Remark~\ref{working} and in the pseudocodes for the algorithms.

\section{Low-rank approximation of arbitrary matrices}
\label{low-rank}

As discussed by~\cite{halko-martinsson-tropp}, randomized algorithms
permit the efficient calculation of nearly optimal rank-$k$ approximations
to a given $m \times n$ matrix $A$, that is, of matrices $U$, $\Sigma$, and $V$
such that $U$ is $m \times k$, $\Sigma$ is $k \times k$, $V$ is $n \times k$,
the columns of $U$ are orthonormal, as are the columns of $V$,
$\Sigma$ is diagonal and its entries are nonnegative, and
\begin{equation}
\| A - U \Sigma V^* \|_2 \approx \sigma_{k+1}(A),
\end{equation}
where $\| A - U \Sigma V^* \|_2$ denotes the spectral norm
of $A - U \Sigma V^*$, and $\sigma_{k+1}(A)$ is the spectral-norm accuracy
of the best approximation to $A$ of rank at most $k$
(which is also equal to the $(k+1)$st greatest singular value of $A$).

Our codes implement Algorithms~4.4 and~5.1 of~\cite{halko-martinsson-tropp},
duplicated here as Algorithms~\ref{subspits} and~\ref{directSVD}, respectively.
The output of Algorithm~\ref{subspits} feeds into Algorithm~\ref{directSVD}.
Algorithm~\ref{subspits} is based on tall-skinny matrix factorizations
of the form $Q \cdot R$,
where the columns of $Q$ are orthonormal and $R$ is square
($R$ need not be triangular, however). Given a matrix factorization
of the form $U \cdot \Sigma \cdot V^*$, where the columns of $U$
are orthonormal, as are the columns of $V$, and where both $\Sigma$ and $V$
are square, we use $Q = U$ and $R = \Sigma V^*$ to obtain a factorization
of the form $Q \cdot R$. In our implementations,
we obtain matrix factorizations of the form $U \cdot \Sigma \cdot V^*$
via the methods evaluated in Section~\ref{tssvd} above.
Below, we compare the results of using the two different methods evaluated
in Section~\ref{tssvd} above for the tall-skinny matrix factorizations
required in Algorithm~\ref{subspits}, always running Algorithm~\ref{subspits}
and feeding its output into Algorithm~\ref{directSVD}.

We run the tall-skinny factorization twice in succession
only for the very last step in Algorithm~\ref{subspits};
the purpose of the earlier steps in Algorithm~\ref{subspits}
is to track a subspace, and so long as the column spaces
of the resulting matrices are accurate,
then whether the columns are numerically orthonormal matters little (in fact,
replacing $Q$ with the lower triangular/trapezoidal factor $L$
in an $LU$ factorization is sufficient, as shown
by~\cite{shabat-shmueli-aizenbud-averbuch}
and~\cite{li-linderman-szlam-stanton-kluger-tygert}
--- the column space of $L$ is the same as the column space of $Q$).

In principle, the last three steps in Algorithm~\ref{tsSVD} are superfluous
for the task of tracking a subspace and could be omitted.
However, in the interest of modular programming,
we feed Algorithm~\ref{subspits} with the final results
of Algorithms~\ref{tsSVD} and~\ref{GramSVD}
rather than with intermediate results; we found the extra costs
to be tolerable, anyways.

The remainder of the present section gives empirical results
on Algorithm~\ref{subspits} feeding into Algorithm~\ref{directSVD},
when using in Algorithm~\ref{subspits}
the two different methods evaluated in Section~\ref{tssvd};
the resulting combinations are Algorithms~\ref{comboa} and~\ref{combob}.
This section also presents empirical results on the existing implementation
of low-rank approximation in Spark's MLlib, which is based
on the implicitly restarted Arnoldi method in ARPACK of~\cite{arpack}.
Our implementations of the algorithms in the present section process
a BlockMatrix from Spark's MLlib (a BlockMatrix can handle matrices
that are not skinny enough for a full row to fit in memory
on a single machine).

Our software includes examples of matrices with many different distributions
of singular values and singular vectors, while for clarity
(as with Section~\ref{tssvd}), the results we present below
pertain to the class of matrices defined in~(\ref{originalmat}) ---
matrices for which the various algorithms produced accuracies near the worst
that we encountered in our experiments.
In~(\ref{originalmat}) for the present section, the only entries
of $\Sigma$ that are potentially nonzero are
\begin{equation}
\label{testSl}
\Sigma_{j,j} = \exp\left( \frac{j-1}{l-1} \cdot \ln\left(10^{-20}\right) \right)
\end{equation}
for $j = 1$, $2$, \dots, $l$. Notice that~(\ref{testSl}) is the same
as~(\ref{testS}) when replacing $l$ in~(\ref{testSl}) with $n$.

Also as in Section~\ref{tssvd},
the headings of the tables have the meanings detailed in Table~\ref{heads}
(which defines $l$). Our Spark environment is detailed in Table~\ref{sparkset}.
We checked accuracies exactly as in Section~\ref{tssvd} and again were sure
to exclude the time spent checking the accuracy from the timings reported
in the tables.

Tables~\ref{partialSVDtime} and~\ref{partialSVDerr} consider sizes of matrices
that are too large for computing all possible singular values
and singular vectors (rather than just a low-rank approximation)
on our cluster with Spark.
Table~\ref{partialSVDtime} indicates that, on our cluster with our version
of Spark, the timings for Algorithm~\ref{comboa} are similar
to the timings for Algorithm~\ref{combob}.
At the same time, Table~\ref{partialSVDerr} indicates that
the reconstruction error $\|A-U \Sigma V^*\|_2$ for Algorithm~\ref{comboa}
is superior to the error for Algorithm~\ref{combob},
while the other notions of accuracy are comparable for both.
Thus, on our cluster with our version of Spark,
Algorithm~\ref{comboa} makes more sense than Algorithm~\ref{combob}.

Tables~\ref{tab56big}, \ref{tab56med}, and~\ref{tab56tiny} correspond
to Tables~\ref{tab1234big}, \ref{tab1234med}, and~\ref{tab1234tiny}.
In accord with Tables~\ref{partialSVDtime} and~\ref{partialSVDerr},
on our cluster with our version of Spark,
the timings for Algorithm~\ref{comboa} are similar
to the timings for Algorithm~\ref{combob},
while the reconstruction error $\|A-U \Sigma V^*\|_2$
for Algorithm~\ref{comboa} is superior to the error for Algorithm~\ref{combob},
and the other notions of accuracy are comparable for both.
Tables~\ref{tab56big}, \ref{tab56med}, and~\ref{tab56tiny} thus
also show that, on our cluster with our version of Spark,
Algorithm~\ref{comboa} makes more sense than Algorithm~\ref{combob}.

The timings in Table~\ref{partialSVDtime} and the errors
in Table~\ref{partialSVDerr} are as expected, as are all the results
in Tables~\ref{tab56big}, \ref{tab56med}, and~\ref{tab56tiny},
with the timings roughly proportional to $l$ times the numbers of entries
in the matrices, and with the errors adhering to the working precision
mentioned in Remark~\ref{working} and in the pseudocodes for the algorithms.

\section{Conclusion}
\label{conclusion}

The numerical experiments reported above illustrate that
the algorithms detailed in this paper outperform (or at least match)
the stock implementations for Spark's MLlib with respect to both accuracy and
efficiency. As Spark's library for machine learning migrates to the upcoming
DataFrame format, it could incorporate these algorithms, as could
other platforms for distributed computation.
The key is attention to details elaborated above,
including randomization, explicit normalization of singular vectors,
and choosing carefully between single and double orthonormalization
and between strict orthonormality and merely tracking subspaces
during various stages of the algorithms.

\begin{acknowledgements}
We would like to thank the anonymous editor and referees
for shaping the presentation.
\end{acknowledgements}

\LinesNumbered
\begin{algorithm}
\caption{Randomized singular value decomposition of tall and skinny matrices
(from~\cite{ballard-demmel-dumitriu})}
\label{tsSVD}
\KwIn{A tall and skinny real matrix $A$}
\KwOut{Real matrices $U$, $\Sigma$, and $V$ such that $A = U \Sigma V^*$,
the columns of $U$ are orthonormal, as are the columns of $V$,
and $\Sigma$ is diagonal and its entries are nonnegative}

Apply an appropriately random orthogonal matrix $\Omega$
(see Remark~\ref{randomness} regarding ``appropriately random'')
to every column of $A^*$, obtaining $B = \Omega A^*$.

Using the TSQR method of~\cite{demmel-grigori-hoemmen-langou},
compute a factorization $B^* = Q R$, where the columns of $Q$ are orthonormal,
and $R$ is upper triangular.

Discard the rows of $R$ corresponding to diagonal entries
which are zero, and discard the corresponding columns of $Q$,
too (if working in finite-precision arithmetic, view any diagonal entry of $R$
as numerically zero that is less than the first diagonal entry
of $R$ times the working precision).

Calculate the singular value decomposition $R = \tilde{U} \Sigma \tilde{V}^*$,
where the columns of $\tilde{U}$ are orthonormal,
as are the columns of $\tilde{V}$,
and $\Sigma$ is diagonal and its entries are nonnegative.

Form $U = Q \tilde{U}$.

Apply the inverse of the random orthogonal matrix $\Omega$ from Step~1
to every column of $\tilde{V}$, obtaining $V = \Omega^{-1} \tilde{V}$
(as $\Omega$ is orthogonal, $\Omega^{-1} = \Omega^*$).
\end{algorithm}

\LinesNumbered
\begin{algorithm}
\caption{Randomized singular value decomposition of tall and skinny matrices
(from~\cite{ballard-demmel-dumitriu}), with double orthonormalization}
\label{tsSVD2}
\KwIn{A tall and skinny real matrix $A$}
\KwOut{Real matrices $U$, $\Sigma$, and $V$ such that $A = U \Sigma V^*$,
the columns of $U$ are orthonormal, as are the columns of $V$,
and $\Sigma$ is diagonal and its entries are nonnegative}

Apply an appropriately random orthogonal matrix $\Omega$
(see Remark~\ref{randomness} regarding ``appropriately random'')
to every column of $A^*$, obtaining $B = \Omega A^*$.

Using the TSQR method of~\cite{demmel-grigori-hoemmen-langou},
compute a factorization $B^* = \tilde{Q} \tilde{R}$,
where the columns of $\tilde{Q}$ are orthonormal,
and $\tilde{R}$ is upper triangular.

Discard the rows of $\tilde{R}$ corresponding to diagonal entries
which are zero, and discard the corresponding columns of $\tilde{Q}$,
too (if working in finite-precision arithmetic, view any diagonal entry
of $\tilde{R}$ as numerically zero that is less than the first diagonal entry
of $\tilde{R}$ times the working precision).

Using the TSQR method of~\cite{demmel-grigori-hoemmen-langou},
compute a factorization $\tilde{Q} = Q R$,
where the columns of $Q$ are orthonormal, and $R$ is upper triangular.

Discard the rows of $R$ corresponding to diagonal entries
which are zero, and discard the corresponding columns of $Q$,
too (if working in finite-precision arithmetic, view any diagonal entry of $R$
as numerically zero that is less than the first diagonal entry
of $R$ times the working precision).

Form $T = R \tilde{R}$.

Calculate the singular value decomposition $T = \tilde{U} \Sigma \tilde{V}^*$,
where the columns of $\tilde{U}$ are orthonormal,
as are the columns of $\tilde{V}$,
and $\Sigma$ is diagonal and its entries are nonnegative.

Form $U = Q \tilde{U}$.

Apply the inverse of the random orthogonal matrix $\Omega$ from Step~1
to every column of $\tilde{V}$, obtaining $V = \Omega^{-1} \tilde{V}$
(as $\Omega$ is orthogonal, $\Omega^{-1} = \Omega^*$).
\end{algorithm}

\LinesNumbered
\begin{algorithm}
\caption{Gram-based singular value decomposition of tall and skinny matrices
(from~\cite{yamazaki-tomov-dongarra2})}
\label{GramSVD}
\KwIn{A tall and skinny real matrix $A$}
\KwOut{Real matrices $U$, $\Sigma$, and $V$ such that $A = U \Sigma V^*$,
the columns of $U$ are orthonormal, as are the columns of $V$,
and $\Sigma$ is diagonal and its entries are nonnegative}

Form the Gram matrix $B = A^* A$.

Calculate the eigendecomposition $B = V D V^*$,
where the columns of $V$ are orthonormal,
and $D$ is diagonal and its entries are nonnegative.

Form $\tilde{U} = A V$.

Set $\Sigma$ to be the diagonal matrix whose diagonal entries
are the Euclidean norms of the columns of $\tilde{U}$,
in accord with Remark~\ref{normalizing}.

Discard the columns and rows of $\Sigma$ corresponding to diagonal entries
which are zero, and discard the corresponding columns of $\tilde{U}$ and $V$,
too (if working in finite-precision arithmetic, view any entry of $\Sigma$
as numerically zero that is less than the greatest entry
of $\Sigma$ times the square root of the working precision).

Form $U = \tilde{U} \Sigma^{-1}$, in accord with Remark~\ref{normalizing}.
\end{algorithm}

\LinesNumbered
\begin{algorithm}
\caption{Gram-based singular value decomposition of tall and skinny matrices
(from~\cite{yamazaki-tomov-dongarra2}), with double orthonormalization}
\label{GramSVD2}
\KwIn{A tall and skinny real matrix $A$}
\KwOut{Real matrices $U$, $\Sigma$, and $V$ such that $A = U \Sigma V^*$,
the columns of $U$ are orthonormal, as are the columns of $V$,
and $\Sigma$ is diagonal and its entries are nonnegative}

Form the Gram matrix $B = A^* A$.

Calculate the eigendecomposition $B = \tilde{V} \tilde{D} \tilde{V}^*$,
where the columns of $\tilde{V}$ are orthonormal,
and $\tilde{D}$ is diagonal and its entries are nonnegative.

Form $\tilde{Y} = A \tilde{V}$.

Set $\tilde{\Sigma}$ to be the diagonal matrix whose diagonal entries
are the Euclidean norms of the columns of $\tilde{Y}$,
in accord with Remark~\ref{normalizing}.

Discard the columns and rows of $\tilde{\Sigma}$ corresponding
to diagonal entries which are zero, and discard the corresponding columns
of $\tilde{Y}$ and $\tilde{V}$, too (if working in finite-precision arithmetic,
view any entry of $\tilde{\Sigma}$ as numerically zero that is less than
the greatest entry of $\tilde{\Sigma}$ times the square root
of the working precision).

Form $Y = \tilde{Y} \tilde{\Sigma}^{-1}$,
in accord with Remark~\ref{normalizing}.

Form the Gram matrix $Z = Y^* Y$.

Calculate the eigendecomposition $Z = W D W^*$,
where the columns of $W$ are orthonormal,
and $D$ is diagonal and its entries are nonnegative.

Form $\tilde{Q} = Y W$.

Set $T$ to be the diagonal matrix whose diagonal entries
are the Euclidean norms of the columns of $\tilde{Q}$,
in accord with Remark~\ref{normalizing}.

Discard the columns and rows of $T$ corresponding to diagonal entries
which are zero, and discard the corresponding columns of $\tilde{Q}$ and $W$,
too (if working in finite-precision arithmetic, view any entry of $T$
as numerically zero that is less than the greatest entry
of $T$ times the square root of the working precision).

Form $Q = \tilde{Q} T^{-1}$, in accord with Remark~\ref{normalizing}.

Form $R = T W^* \tilde{\Sigma} \tilde{V}^*$.

Calculate the singular value decomposition $R = P \Sigma V^*$,
where the columns of $P$ are orthonormal, as are the columns of $V$,
and $\Sigma$ is diagonal and its entries are nonnegative.

Form $U = Q P$.
\end{algorithm}

\LinesNumbered
\begin{algorithm}
\caption{Randomized subspace iteration
(Algorithm~4.4 of~\cite{halko-martinsson-tropp})}
\label{subspits}
\KwIn{A real $m \times n$ matrix $A$ and integers $l$ and $i$
such that $0 < l < \min(m,n)$ and $i \ge 0$; the number of iterations is $i$}
\KwOut{A real $m \times l$ matrix $Q$ whose columns are orthonormal
and whose range approximates the range of $A$, in the sense that
the spectral norm $\| A - Q Q^* A \|_2$ is small}

Form an $n \times l$ matrix $\tilde{Q}_0$
whose entries are independent and identically distributed
centered Gaussian random variables.

\For{$j = 1$ \KwTo $i$}{
Form $Y_j = A \tilde{Q}_{j-1}$.

Compute a factorization $Y_j = Q_j R_j$,
where the columns of $Q_j$ are orthonormal and $R_j$ is square,
using Algorithm~\ref{tsSVD} or Algorithm~\ref{GramSVD}
(as described at the beginning of Section~\ref{low-rank}).

Form $\tilde{Y}_j = A^* Q_j$.

Compute a factorization $\tilde{Y}_j = \tilde{Q}_j \tilde{R}_j$,
where the columns of $\tilde{Q}_j$ are orthonormal
and $\tilde{R}_j$ is square,
using Algorithm~\ref{tsSVD} or Algorithm~\ref{GramSVD}
(as described at the beginning of Section~\ref{low-rank}).
}

Form $Y = A \tilde{Q}_i$.

Compute a factorization $Y = Q R$,
where the columns of $Q$ are orthonormal and $R$ is square,
using in this last step the double orthonormalization
of Algorithms~\ref{tsSVD2} and~\ref{GramSVD2},
not the single orthonormalization
of Algorithms~\ref{tsSVD} and~\ref{GramSVD}
(and, again, see the beginning of Section~\ref{low-rank}).
\end{algorithm}

\LinesNumbered
\begin{algorithm}
\caption{Straightforward singular value decomposition
(Algorithm~5.1 of~\cite{halko-martinsson-tropp})}
\label{directSVD}
\KwIn{Matrices $A$ and $Q$ such that the spectral norm $\| A - Q Q^* A \|_2$
is small and the columns of $Q$ are orthonormal (the matrix $Q$ output
from Algorithm~\ref{subspits} is suitable for the input here)}
\KwOut{Matrices $U$, $\Sigma$, and $V$ such that the spectral norm
$\| A - U \Sigma V^* \|_2$ is small, the columns of $U$ are orthonormal,
as are the columns of $V$, and $\Sigma$ is diagonal
and its entries are nonnegative}

Form the matrix $B = Q^* A$.

Compute a singular value decomposition,
$B = \tilde{U} \Sigma V^*$, where the columns of $\tilde{U}$ are orthonormal,
as are the columns of $V$, and $\Sigma$ is diagonal
and its entries are nonnegative.

Form the matrix $U = Q \tilde{U}$.

\end{algorithm}

\clearpage

\LinesNumbered
\begin{algorithm}
\caption{Algorithm~\ref{directSVD} fed with the results
of Algorithm~\ref{subspits} using Algorithms~\ref{tsSVD} and~\ref{tsSVD2}}
\label{comboa}
\KwIn{A real $m \times n$ matrix $A$ and integers $l$ and $i$
such that $0 < l < \min(m,n)$ and $i \ge 0$; the number of iterations
for Algorithm~\ref{subspits} is $i$}
\KwOut{Matrices $U$, $\Sigma$, and $V$ such that the spectral norm
$\| A - U \Sigma V^* \|_2$ is small, the columns of $U$ are orthonormal,
as are the columns of $V$, and $\Sigma$ is diagonal
and its entries are nonnegative}

Compute a real $m \times l$ matrix $Q$ whose columns are orthonormal,
such that the spectral norm $\| A - Q Q^* A \|_2$ is small,
via Algorithm~\ref{subspits} with $i$ iterations, using Algorithm~\ref{tsSVD}
in Algorithm~\ref{subspits}'s Steps~4 and~6 and Algorithm~\ref{tsSVD2}
in Algorithm~\ref{subspits}'s last step.

Compute matrices $U$, $\Sigma$, and $V$ such that the spectral norm
$\| A - U \Sigma V^* \|_2$ is small, the columns of $U$ are orthonormal,
as are the columns of $V$, and $\Sigma$ is diagonal
and its entries are nonnegative, via Algorithm~\ref{directSVD}
fed with matrices $A$ and $Q$ from the first step of the present algorithm.

\end{algorithm}

\LinesNumbered
\begin{algorithm}
\caption{Algorithm~\ref{directSVD} fed with the results
of Algorithm~\ref{subspits} using Algorithms~\ref{GramSVD} and~\ref{GramSVD2}}
\label{combob}
\KwIn{A real $m \times n$ matrix $A$ and integers $l$ and $i$
such that $0 < l < \min(m,n)$ and $i \ge 0$; the number of iterations
for Algorithm~\ref{subspits} is $i$}
\KwOut{Matrices $U$, $\Sigma$, and $V$ such that the spectral norm
$\| A - U \Sigma V^* \|_2$ is small, the columns of $U$ are orthonormal,
as are the columns of $V$, and $\Sigma$ is diagonal
and its entries are nonnegative}

Compute a real $m \times l$ matrix $Q$ whose columns are orthonormal,
such that the spectral norm $\| A - Q Q^* A \|_2$ is small,
via Algorithm~\ref{subspits} with $i$ iterations, using Algorithm~\ref{GramSVD}
in Algorithm~\ref{subspits}'s Steps~4 and~6 and Algorithm~\ref{GramSVD2}
in Algorithm~\ref{subspits}'s last step.

Compute matrices $U$, $\Sigma$, and $V$ such that the spectral norm
$\| A - U \Sigma V^* \|_2$ is small, the columns of $U$ are orthonormal,
as are the columns of $V$, and $\Sigma$ is diagonal
and its entries are nonnegative, via Algorithm~\ref{directSVD}
fed with matrices $A$ and $Q$ from the first step of the present algorithm.

\end{algorithm}

\begin{table}
\caption{Meanings of the headings in the other tables}
\label{heads}
{
\begin{tabular}{rp{3.3in}}
\hline\noalign{\smallskip}
Heading & Meaning \\
\noalign{\smallskip}\hline\noalign{\smallskip}
$m$ & number of rows in the matrix being decomposed or approximated \\
$n$ & number of columns in the matrix being decomposed or approximated \\
$l$ & rank of the approximation being constructed in Algorithms~\ref{comboa}
and~\ref{combob} (for the tables using Algorithms~\ref{comboa}
and~\ref{combob}) \\
$i$ & number of iterations used in Algorithm~\ref{subspits} (for the tables
using Algorithms~\ref{comboa} and~\ref{combob}, both of which leverage
Algorithm~\ref{subspits}) \\
Algorithm & specifies the number of the algorithm used
(or ``pre-existing'' for the original implementation in Spark) \\
CPU Time & sum over all CPU cores in all executors of the time in seconds
spent actually processing \\
Wall-Clock & sum over all executors of the time in seconds that they were
reserved \\
$\| A - U \Sigma V^* \|_2$ & spectral norm of the discrepancy between the
computed approximation $U \Sigma V^*$ and the matrix $A$ being decomposed
or approximated; {\it please note that our setting for the  working precision
largely determines this error --- see Remark~\ref{working} and the steps
in the algorithms, ``Discard\dots.''} \\
$\entrywisemax(|U^* U - I|)$ & maximal absolute value of the entries
in the difference between $U^* U$ and the identity matrix $I$,
where $U \Sigma V^*$ is the computed approximation \\
$\entrywisemax(|V^* V - I|)$ & maximal absolute value of the entries
in the difference between $V^* V$ and the identity matrix $I$,
where $U \Sigma V^*$ is the computed approximation \\
\noalign{\smallskip}\hline
\end{tabular}
}
\end{table}

\begin{table}
\caption{Settings for Spark}
\label{sparkset}
{
\small
\begin{tabular}{ll}
\hline\noalign{\smallskip}
Parameter & Setting \\
\noalign{\smallskip}\hline\noalign{\smallskip}
spark.dynamicAllocation.maxExecutors & 180 \\
spark.executor.cores & 30 \\
spark.executor.memory & 60g \\
rowsPerPart (in a BlockMatrix)$^\dagger$ & 1024 \\
colsPerPart (in a BlockMatrix) & 1024 \\
Spark version & 2.0.1 \\
total machines available & 200 \\
BLAS-LAPACK library & Intel MKL \\
\noalign{\smallskip}\hline
\end{tabular}

\parbox{3.25in}{\ }

\parbox{3.25in}{\ }

\parbox{2.92in}{
$^\dagger$This is also the number of rows in a block
of the resilient distributed dataset that underlies
an IndexedRowMatrix. Our software converts the matrix
in formula~(\ref{originalmat})
from a BlockMatrix to an IndexedRowMatrix whenever necessary,
which preserves the number of rows per block.
}
}
\end{table}

\clearpage

\begin{table}
\caption{$m =$ 1,000,000; $n =$ 2,000}
\label{tab1234big}
{
\small
\begin{tabular}{rccccc}
\hline\noalign{\smallskip}
&&&& $\entrywisemax($ & $\entrywisemax($ \\
Algorithm & CPU Time & Wall-Clock & $\|A-U \Sigma V^*\|_2$ &
$\phantom{(}|U^*U-I|)$ & $\phantom{(}|V^*V-I|)$ \\
\noalign{\smallskip}\hline\noalign{\smallskip}
1 & 1.48E+04 & 1.48E+04 & 9.76E-12 & 6.84E-06 & 3.51E-15 \\
2 & 6.84E+04 & 9.01E+04 & 9.76E-12 & 6.44E-13 & 4.68E-15 \\
3 & 1.33E+04 & 1.67E+04 & 9.92E-08 & 6.20E-04 & 1.73E-14 \\
4 & 1.36E+04 & 2.52E+04 & 9.64E-07 & 1.10E-14 & 2.90E-15 \\
pre-existing & 1.12E+04 & 1.28E+04 & 1.83E-09 & 2.34E-00 & 3.12E-15 \\
\noalign{\smallskip}\hline
\end{tabular}
}
\end{table}

\begin{table}
\caption{$m =$ 100,000; $n =$ 2,000}
\label{tab1234med}
{
\small
\begin{tabular}{rccccc}
\hline\noalign{\smallskip}
&&&& $\entrywisemax($ & $\entrywisemax($ \\
Algorithm & CPU Time & Wall-Clock & $\|A-U \Sigma V^*\|_2$ &
$\phantom{(}|U^*U-I|)$ & $\phantom{(}|V^*V-I|)$ \\
\noalign{\smallskip}\hline\noalign{\smallskip}
1 & 1.59E+03 & 1.02E+03 & 9.76E-12 & 5.47E-06 & 3.22E-15 \\
2 & 6.85E+03 & 3.39E+03 & 9.76E-12 & 6.85E-13 & 4.06E-15 \\
3 & 1.32E+03 & 9.19E+02 & 9.92E-08 & 3.11E-04 & 1.22E-14 \\
4 & 1.58E+03 & 1.30E+03 & 9.64E-07 & 6.66E-15 & 2.69E-15 \\
pre-existing & 1.27E+03 & 9.68E+02 & 2.75E-15 & 9.91E-01 & 2.50E-15 \\
\noalign{\smallskip}\hline
\end{tabular}
}
\end{table}

\begin{table}
\caption{$m =$ 10,000; $n =$ 2,000}
\label{tab1234tiny}
{
\small
\begin{tabular}{rccccc}
\hline\noalign{\smallskip}
&&&& $\entrywisemax($ & $\entrywisemax($ \\
Algorithm & CPU Time & Wall-Clock & $\|A-U \Sigma V^*\|_2$ &
$\phantom{(}|U^*U-I|)$ & $\phantom{(}|V^*V-I|)$ \\
\noalign{\smallskip}\hline\noalign{\smallskip}
1 & 3.86E+02 & 8.40E+01 & 9.76E-12 & 4.35E-06 & 3.55E-15 \\
2 & 9.26E+02 & 1.42E+02 & 9.76E-12 & 7.67E-12 & 3.19E-15 \\
3 & 2.52E+02 & 5.60E+01 & 9.92E-08 & 2.15E-04 & 1.82E-14 \\
4 & 3.16E+02 & 8.40E+01 & 9.64E-07 & 6.66E-15 & 3.33E-15 \\
pre-existing & 2.15E+02 & 7.30E+01 & 1.89E-15 & 9.97E-01 & 2.57E-15 \\
\noalign{\smallskip}\hline
\end{tabular}
}
\end{table}

\clearpage

\begin{table}
\caption{$m =$ 1,000,000; $n =$ 2,000; $l = $ 20; $i = $ 2}
\label{tab56big}
{
\small
\begin{tabular}{rccccc}
\hline\noalign{\smallskip}
&&&& $\entrywisemax($ & $\entrywisemax($ \\
Algorithm & CPU Time & Wall-Clock & $\|A-U \Sigma V^*\|_2$ &
$\phantom{(}|U^*U-I|)$ & $\phantom{(}|V^*V-I|)$ \\
\noalign{\smallskip}\hline\noalign{\smallskip}
7 & 3.06E+03 & 8.80E+03 & 2.64E-12 & 4.44E-15 & 8.88E-16 \\
8 & 2.80E+03 & 9.94E+03 & 4.83E-07 & 3.77E-15 & 5.55E-16 \\
pre-existing & 6.06E+03 & 1.16E+04 & 3.36E-10 & 1.00E-00 & 6.66E-16 \\
\noalign{\smallskip}\hline
\end{tabular}
}
\end{table}

\begin{table}
\caption{$m =$ 100,000; $n =$ 2,000; $l = $ 20; $i = $ 2}
\label{tab56med}
{
\small
\begin{tabular}{rccccc}
\hline\noalign{\smallskip}
&&&& $\entrywisemax($ & $\entrywisemax($ \\
Algorithm & CPU Time & Wall-Clock & $\|A-U \Sigma V^*\|_2$ &
$\phantom{(}|U^*U-I|)$ & $\phantom{(}|V^*V-I|)$ \\
\noalign{\smallskip}\hline\noalign{\smallskip}
7 & 3.28E+02 & 4.78E+02 & 2.64E-12 & 3.11E-15 & 1.44E-15 \\
8 & 4.33E+02 & 4.71E+02 & 4.83E-07 & 1.55E-15 & 8.36E-16 \\
pre-existing & 6.17E+02 & 4.92E+02 & 3.36E-10 & 1.00E-00 & 4.44E-16 \\
\noalign{\smallskip}\hline
\end{tabular}
}
\end{table}

\begin{table}
\caption{$m =$ 10,000; $n =$ 2,000; $l = $ 20; $i = $ 2}
\label{tab56tiny}
{
\small
\begin{tabular}{rccccc}
\hline\noalign{\smallskip}
&&&& $\entrywisemax($ & $\entrywisemax($ \\
Algorithm & CPU Time & Wall-Clock & $\|A-U \Sigma V^*\|_2$ &
$\phantom{(}|U^*U-I|)$ & $\phantom{(}|V^*V-I|)$ \\
\noalign{\smallskip}\hline\noalign{\smallskip}
7 & 7.20E+01 & 7.50E+01 & 2.64E-12 & 2.22E-15 & 1.89E-15 \\
8 & 8.00E+01 & 9.30E+01 & 4.83E-07 & 6.66E-16 & 6.66E-16 \\
pre-existing & 1.18E+02 & 9.40E+01 & 3.36E-10 & 1.00E-00 & 6.66E-16 \\
\noalign{\smallskip}\hline
\end{tabular}
}
\end{table}

\begin{table}
\caption{Timings for $l = $ 10; $i = $ 2}
\label{partialSVDtime}
{
\small
\begin{tabular}{rrrcc}
\hline\noalign{\smallskip}
Algorithm & $m$ & $n$ & CPU Time & Wall-Clock \\
\noalign{\smallskip}\hline\noalign{\smallskip}
7 & 100,000 & 100,000 & 1.04E+04 & 4.88E+03 \\
8 & 100,000 & 100,000 & 9.52E+03 & 7.41E+03 \\
\noalign{\smallskip}\hline\noalign{\smallskip}
7 & 1,000,000 & 10,000 & 9.11E+03 & 1.05E+04 \\
8 & 1,000,000 & 10,000 & 9.56E+03 & 1.01E+04 \\
\noalign{\smallskip}\hline\noalign{\smallskip}
7 & 100,000 & 10,000 & 1.10E+03 & 5.40E+02 \\
8 & 100,000 & 10,000 & 1.02E+03 & 4.93E+02 \\
\noalign{\smallskip}\hline
\end{tabular}
}
\end{table}

\begin{table}
\caption{Errors for $l = $ 10; $i = $ 2}
\label{partialSVDerr}
{
\small
\begin{tabular}{rrrccc}
\hline\noalign{\smallskip}
&&&& $\entrywisemax($ & $\entrywisemax($ \\
Algorithm & $m$ & $n$ & $\|A-U \Sigma V^*\|_2$ &
$\phantom{(}|U^*U-I|)$ & $\phantom{(}|V^*V-I|)$ \\
\noalign{\smallskip}\hline\noalign{\smallskip}
7 & 100,000 & 100,000 & 7.74E-12 & 6.66E-16 & 1.78E-15 \\
8 & 100,000 & 100,000 & 2.15E-07 & 7.77E-16 & 1.33E-15 \\
\noalign{\smallskip}\hline\noalign{\smallskip}
7 & 1,000,000 & 10,000 & 7.74E-12 & 3.00E-15 & 7.77E-16 \\
8 & 1,000,000 & 10,000 & 2.15E-07 & 2.89E-15 & 4.44E-16 \\
\noalign{\smallskip}\hline\noalign{\smallskip}
7 & 100,000 & 10,000 & 7.74E-12 & 1.22E-15 & 9.99E-16 \\
8 & 100,000 & 10,000 & 2.15E-07 & 2.86E-16 & 4.44E-16 \\
\noalign{\smallskip}\hline
\end{tabular}
}
\end{table}

\clearpage

\appendix
\normalsize
\section{Restricting to ten times fewer executors}
\label{only18}

Tables~\ref{tab1234big18}--\ref{tab1234tiny18},
Tables~\ref{tab56big18}--\ref{tab56tiny18},
and Tables~\ref{partialSVDtime18} and~\ref{partialSVDerr18} display results
analogous to those in Tables~\ref{tab1234big}--\ref{tab1234tiny},
Tables~\ref{tab56big}--\ref{tab56tiny},
and Tables~\ref{partialSVDtime} and~\ref{partialSVDerr},
but with the number of executors, spark.dynamicAllocation.maxExecutors,
set to 18 (rather than 180).
The results are broadly comparable to those presented earlier.
This indicates how the timings scale with the number of machines.
Of course, other processing in Spark (not necessarily related
to principal component analysis or singular value decomposition) can benefit
from having the data stored over more executors, and moving data around
the cluster can dominate the overall timings in real-world usage
(see also Remark~\ref{goal} in the introduction of the present paper).

\begin{table}
\caption{$m =$ 1,000,000; $n =$ 2,000; restricted to ten times fewer executors}
\label{tab1234big18}
{
\small
\begin{tabular}{rccccc}
\hline\noalign{\smallskip}
&&&& $\entrywisemax($ & $\entrywisemax($ \\
Algorithm & CPU Time & Wall-Clock & $\|A-U \Sigma V^*\|_2$ &
$\phantom{(}|U^*U-I|)$ & $\phantom{(}|V^*V-I|)$ \\
\noalign{\smallskip}\hline\noalign{\smallskip}
1 & 9.23E+03 & 4.72E+03 & 9.76E-12 & 6.21E-06 & 3.00E-15 \\
2 & 5.91E+04 & 5.44E+04 & 9.76E-12 & 6.75E-13 & 3.06E-15 \\
3 & 7.36E+03 & 4.14E+03 & 9.92E-08 & 6.13E-04 & 1.38E-14 \\
4 & 1.00E+04 & 7.72E+03 & 9.64E-07 & 1.02E-14 & 2.69E-15 \\
pre-existing & 6.54E+03 & 3.56E+03 & 1.79E-09 & 3.17E-00 & 3.96E-15 \\
\noalign{\smallskip}\hline
\end{tabular}
}
\end{table}

\begin{table}
\caption{$m =$ 100,000; $n =$ 2,000; restricted to ten times fewer executors}
\label{tab1234med18}
{
\small
\begin{tabular}{rccccc}
\hline\noalign{\smallskip}
&&&& $\entrywisemax($ & $\entrywisemax($ \\
Algorithm & CPU Time & Wall-Clock & $\|A-U \Sigma V^*\|_2$ &
$\phantom{(}|U^*U-I|)$ & $\phantom{(}|V^*V-I|)$ \\
\noalign{\smallskip}\hline\noalign{\smallskip}
1 & 1.74E+03 & 8.76E+02 & 9.76E-12 & 5.30E-06 & 3.33E-15 \\
2 & 7.08E+03 & 3.74E+03 & 9.76E-12 & 4.93E-13 & 3.89E-15 \\
3 & 1.26E+03 & 7.36E+02 & 9.92E-08 & 2.33E-04 & 1.87E-14 \\
4 & 1.62E+03 & 1.01E+03 & 9.64E-07 & 5.33E-15 & 3.33E-15 \\
pre-existing & 1.27E+03 & 8.13E+02 & 2.15E-15 & 9.92E-01 & 2.32E-15 \\
\noalign{\smallskip}\hline
\end{tabular}
}
\end{table}

\begin{table}
\caption{$m =$ 10,000; $n =$ 2,000; restricted to ten times fewer executors}
\label{tab1234tiny18}
{
\small
\begin{tabular}{rccccc}
\hline\noalign{\smallskip}
&&&& $\entrywisemax($ & $\entrywisemax($ \\
Algorithm & CPU Time & Wall-Clock & $\|A-U \Sigma V^*\|_2$ &
$\phantom{(}|U^*U-I|)$ & $\phantom{(}|V^*V-I|)$ \\
\noalign{\smallskip}\hline\noalign{\smallskip}
1 & 4.02E+02 & 9.80E+01 & 9.76E-12 & 5.80E-06 & 3.67E-15 \\
2 & 8.69E+02 & 1.70E+02 & 9.76E-12 & 2.65E-11 & 3.88E-15 \\
3 & 2.04E+02 & 6.70E+01 & 9.92E-08 & 2.55E-04 & 1.73E-14 \\
4 & 2.26E+02 & 9.00E+01 & 9.64E-07 & 5.33E-15 & 2.89E-15 \\
pre-existing & 1.86E+02 & 9.50E+01 & 2.45E-15 & 9.96E-01 & 2.36E-15 \\
\noalign{\smallskip}\hline
\end{tabular}
}
\end{table}

\clearpage

\begin{table}
\caption{$m =$ 1,000,000; $n =$ 2,000; $l = $ 20; $i = $ 2;
restricted to ten times fewer executors}
\label{tab56big18}
{
\small
\begin{tabular}{rccccc}
\hline\noalign{\smallskip}
&&&& $\entrywisemax($ & $\entrywisemax($ \\
Algorithm & CPU Time & Wall-Clock & $\|A-U \Sigma V^*\|_2$ &
$\phantom{(}|U^*U-I|)$ & $\phantom{(}|V^*V-I|)$ \\
\noalign{\smallskip}\hline\noalign{\smallskip}
7 & 2.48E+03 & 4.44E+03 & 2.64E-12 & 4.88E-15 & 1.22E-15 \\
8 & 2.33E+03 & 4.47E+03 & 4.83E-07 & 3.33E-15 & 6.66E-16 \\
pre-existing & 5.56E+03 & 6.84E+03 & 3.36E-10 & 1.00E-00 & 6.66E-16 \\
\noalign{\smallskip}\hline
\end{tabular}
}
\end{table}

\begin{table}
\caption{$m =$ 100,000; $n =$ 2,000; $l = $ 20; $i = $ 2;
restricted to ten times fewer executors}
\label{tab56med18}
{
\small
\begin{tabular}{rccccc}
\hline\noalign{\smallskip}
&&&& $\entrywisemax($ & $\entrywisemax($ \\
Algorithm & CPU Time & Wall-Clock & $\|A-U \Sigma V^*\|_2$ &
$\phantom{(}|U^*U-I|)$ & $\phantom{(}|V^*V-I|)$ \\
\noalign{\smallskip}\hline\noalign{\smallskip}
7 & 3.99E+02 & 4.10E+02 & 2.64E-12 & 2.89E-15 & 1.55E-15 \\
8 & 3.28E+02 & 4.05E+02 & 4.83E-07 & 2.44E-15 & 8.88E-16 \\
pre-existing & 6.31E+02 & 5.17E+02 & 3.36E-10 & 1.00E-00 & 8.88E-16 \\
\noalign{\smallskip}\hline
\end{tabular}
}
\end{table}

\begin{table}
\caption{$m =$ 10,000; $n =$ 2,000; $l = $ 20; $i = $ 2;
restricted to ten times fewer executors}
\label{tab56tiny18}
{
\small
\begin{tabular}{rccccc}
\hline\noalign{\smallskip}
&&&& $\entrywisemax($ & $\entrywisemax($ \\
Algorithm & CPU Time & Wall-Clock & $\|A-U \Sigma V^*\|_2$ &
$\phantom{(}|U^*U-I|)$ & $\phantom{(}|V^*V-I|)$ \\
\noalign{\smallskip}\hline\noalign{\smallskip}
7 & 7.60E+01 & 9.80E+01 & 2.64E-12 & 2.66E-15 & 1.55E-15 \\
8 & 6.30E+01 & 7.40E+01 & 4.83E-07 & 2.22E-15 & 1.55E-15 \\
pre-existing & 1.21E+02 & 9.80E+01 & 3.36E-10 & 1.00E-00 & 6.66E-16 \\
\noalign{\smallskip}\hline
\end{tabular}
}
\end{table}

\begin{table}
\caption{Timings for $l = $ 10; $i = $ 2;
restricted to ten times fewer executors}
\label{partialSVDtime18}
{
\small
\begin{tabular}{rrrcc}
\hline\noalign{\smallskip}
Algorithm & $m$ & $n$ & CPU Time & Wall-Clock \\
\noalign{\smallskip}\hline\noalign{\smallskip}
7 & 100,000 & 100,000 & 1.04E+04 & 6.07E+03 \\
8 & 100,000 & 100,000 & 1.02E+04 & 6.28E+03 \\
\noalign{\smallskip}\hline\noalign{\smallskip}
7 & 1,000,000 & 10,000 & 9.36E+03 & 5.93E+03 \\
8 & 1,000,000 & 10,000 & 9.38E+03 & 6.77E+03 \\
\noalign{\smallskip}\hline\noalign{\smallskip}
7 & 100,000 & 10,000 & 1.01E+03 & 5.19E+02 \\
8 & 100,000 & 10,000 & 1.01E+03 & 5.04E+02 \\
\noalign{\smallskip}\hline
\end{tabular}
}
\end{table}

\begin{table}
\caption{Errors for $l = $ 10; $i = $ 2;
restricted to ten times fewer executors}
\label{partialSVDerr18}
{
\small
\begin{tabular}{rrrccc}
\hline\noalign{\smallskip}
&&&& $\entrywisemax($ & $\entrywisemax($ \\
Algorithm & $m$ & $n$ & $\|A-U \Sigma V^*\|_2$ &
$\phantom{(}|U^*U-I|)$ & $\phantom{(}|V^*V-I|)$ \\
\noalign{\smallskip}\hline\noalign{\smallskip}
7 & 100,000 & 100,000 & 7.74E-12 & 1.55E-15 & 1.78E-15 \\
8 & 100,000 & 100,000 & 2.15E-07 & 8.88E-16 & 1.78E-15 \\
\noalign{\smallskip}\hline\noalign{\smallskip}
7 & 1,000,000 & 10,000 & 7.74E-12 & 1.55E-15 & 6.66E-16 \\
8 & 1,000,000 & 10,000 & 2.15E-07 & 1.11E-15 & 1.11E-16 \\
\noalign{\smallskip}\hline\noalign{\smallskip}
7 & 100,000 & 10,000 & 7.74E-12 & 2.00E-15 & 8.88E-16 \\
8 & 100,000 & 10,000 & 2.15E-07 & 8.88E-16 & 7.77E-16 \\
\noalign{\smallskip}\hline
\end{tabular}
}
\end{table}

\clearpage

\section{Another example with ten times fewer executors}
\label{devil}

\begin{figure}
\centering
\includegraphics[width=.5\textwidth]{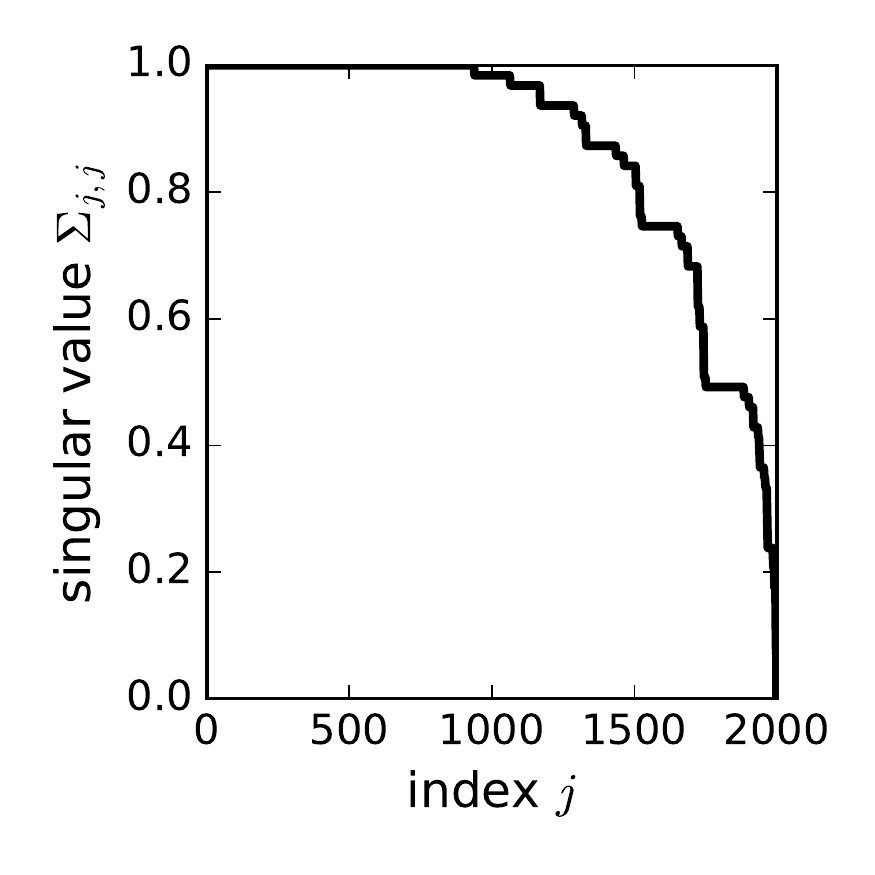}
\caption{Singular values
$\Sigma_{1,1}$, $\Sigma_{2,2}$, \dots, $\Sigma_{2000,2000}$,
which are the diagonal entries of $\Sigma$ in~(\ref{originalmat}),
when $k = n =$ 2,000, for Tables~\ref{tab1234big18new}--\ref{tab1234tiny18new}
in Appendix~\ref{devil}}
\label{singvals}
\end{figure}

Similar to Appendix~\ref{only18}, the present appendix presents
Tables~\ref{tab1234big18new}--\ref{tab1234tiny18new},
Tables~\ref{tab56big18new}--\ref{tab56tiny18new},
and Tables~\ref{partialSVDtime18new} and~\ref{partialSVDerr18new},
reporting results
analogous to those in Tables~\ref{tab1234big}--\ref{tab1234tiny},
Tables~\ref{tab56big}--\ref{tab56tiny},
and Tables~\ref{partialSVDtime} and~\ref{partialSVDerr},
with the same setting as in Appendix~\ref{only18} of the number of executors,
spark.dynamicAllocation.maxExecutors, being 18 (rather than 180).
The present appendix follows an anonymous reviewer's suggestion,
using for the diagonal entries of $\Sigma$ in~(\ref{originalmat})
singular values $\Sigma_{j,j}$
from a fractal ``Devil's staircase'' with many repeated singular values
of varying multiplicities; Figure~\ref{singvals} plots the singular values
for Tables~\ref{tab1234big18new}--\ref{tab1234tiny18new}.
Specifically, the singular values arise from the following Scala code:

\begin{center}
\parbox{.67\textwidth}{
\parindent=0pt
\tt

(0 until k).toArray.map( j =>

\quad Integer.parseInt(Integer.toOctalString(

\quad\quad Math.round(j * Math.pow(8, 6).toFloat / k)

\quad ).replaceAll("[1-7]", "1"), 2)

\quad / Math.pow(2, 6) / (1 - Math.pow(2, -6))

).sorted.reverse

}
\end{center}

\noindent Here, $k = n$
for Tables~\ref{tab1234big18new}--\ref{tab1234tiny18new}
and $k = l$ for Tables~\ref{tab56big18new}--\ref{partialSVDerr18new}.
Thus, the singular values arise from replacing the octal digits 1--7
with the binary digit 1 (keeping the octal digit 0 as the binary digit 0)
for rounded representations of the real numbers between 0 and 1,
then rescaling so that the final singular values range from 0 to 1, inclusive.

Again, the results are broadly comparable to those presented earlier;
in some cases some of the algorithms attain better accuracy on the examples
of the present appendix, but otherwise the numbers in the tables are similar.

\clearpage

\begin{table}
\caption{$m =$ 1,000,000; $n =$ 2,000; restricted to ten times fewer executors;
Appendix~\ref{devil} defines the singular values of the matrix being processed}
\label{tab1234big18new}
{
\small
\begin{tabular}{rccccc}
\hline\noalign{\smallskip}
&&&& $\entrywisemax($ & $\entrywisemax($ \\
Algorithm & CPU Time & Wall-Clock & $\|A-U \Sigma V^*\|_2$ &
$\phantom{(}|U^*U-I|)$ & $\phantom{(}|V^*V-I|)$ \\
\noalign{\smallskip}\hline\noalign{\smallskip}
1 & 9.47E+03 & 1.14E+04 & 1.67E-14 & 6.22E-15 & 3.33E-15 \\
2 & 1.06E+05 & 1.07E+05 & 1.61E-14 & 6.88E-15 & 3.22E-15 \\
3 & 8.91E+03 & 7.65E+03 & 1.84E-14 & 9.24E-14 & 1.78E-14 \\
4 & 3.20E+04 & 3.88E+04 & 2.34E-14 & 8.88E-15 & 3.60E-15 \\
pre-existing & 5.98E+03 & 6.80E+03 & 7.72E-15 & 1.00E-00 & 6.18E-15 \\
\hline
\end{tabular}
}
\end{table}

\begin{table}
\caption{$m =$ 100,000; $n =$ 2,000; restricted to ten times fewer executors;
Appendix~\ref{devil} defines the singular values of the matrix being processed}
\label{tab1234med18new}
{
\small
\begin{tabular}{rccccc}
\hline\noalign{\smallskip}
&&&& $\entrywisemax($ & $\entrywisemax($ \\
Algorithm & CPU Time & Wall-Clock & $\|A-U \Sigma V^*\|_2$ &
$\phantom{(}|U^*U-I|)$ & $\phantom{(}|V^*V-I|)$ \\
\noalign{\smallskip}\hline\noalign{\smallskip}
1 & 1.71E+03 & 8.81E+02 & 1.62E-14 & 4.24E-15 & 3.81E-15 \\
2 & 1.15E+04 & 5.52E+03 & 1.61E-14 & 3.64E-15 & 3.33E-15 \\
3 & 1.58E+03 & 9.55E+02 & 2.27E-14 & 1.46E-13 & 1.85E-14 \\
4 & 4.02E+03 & 2.49E+03 & 2.48E-14 & 4.66E-15 & 4.04E-15 \\
pre-existing & 1.19E+03 & 7.58E+02 & 7.47E-15 & 1.00E-00 & 5.47E-15 \\
\noalign{\smallskip}\hline
\end{tabular}
}
\end{table}

\begin{table}
\caption{$m =$ 10,000; $n =$ 2,000; restricted to ten times fewer executors;
Appendix~\ref{devil} defines the singular values of the matrix being processed}
\label{tab1234tiny18new}
{
\small
\begin{tabular}{rccccc}
\hline\noalign{\smallskip}
&&&& $\entrywisemax($ & $\entrywisemax($ \\
Algorithm & CPU Time & Wall-Clock & $\|A-U \Sigma V^*\|_2$ &
$\phantom{(}|U^*U-I|)$ & $\phantom{(}|V^*V-I|)$ \\
\noalign{\smallskip}\hline\noalign{\smallskip}
1 & 3.35E+02 & 8.30E+01 & 1.70E-14 & 4.97E-15 & 4.95E-15 \\
2 & 1.74E+03 & 1.79E+02 & 1.67E-14 & 4.52E-15 & 5.01E-15 \\
3 & 2.45E+02 & 9.80E+01 & 1.83E-14 & 1.51E-13 & 1.53E-14 \\
4 & 5.96E+02 & 1.30E+02 & 2.36E-14 & 5.23E-15 & 4.94E-15 \\
pre-existing & 2.11E+02 & 8.40E+01 & 6.23E-15 & 1.00E-00 & 3.82E-15 \\
\noalign{\smallskip}\hline
\end{tabular}
}
\end{table}

\clearpage

\begin{table}
\caption{$m =$ 1,000,000; $n =$ 2,000; $l = $ 20; $i = $ 2;
restricted to ten times fewer executors;
Appendix~\ref{devil} defines the singular values of the matrix being processed}
\label{tab56big18new}
{
\small
\begin{tabular}{rccccc}
\hline\noalign{\smallskip}
&&&& $\entrywisemax($ & $\entrywisemax($ \\
Algorithm & CPU Time & Wall-Clock & $\|A-U \Sigma V^*\|_2$ &
$\phantom{(}|U^*U-I|)$ & $\phantom{(}|V^*V-I|)$ \\
\noalign{\smallskip}\hline\noalign{\smallskip}
7 & 3.49E+03 & 1.09E+04 & 2.69E-15 & 2.00E-15 & 1.55E-15 \\
8 & 3.20E+03 & 1.11E+04 & 8.65E-15 & 3.44E-15 & 8.88E-16 \\
pre-existing & 6.34E+03 & 1.96E+04 & 2.12E-15 & 1.00E-00 & 6.66E-16 \\
\hline
\end{tabular}
}
\end{table}

\begin{table}
\caption{$m =$ 100,000; $n =$ 2,000; $l = $ 20; $i = $ 2;
restricted to ten times fewer executors;
Appendix~\ref{devil} defines the singular values of the matrix being processed}
\label{tab56med18new}
{
\small
\begin{tabular}{rccccc}
\hline\noalign{\smallskip}
&&&& $\entrywisemax($ & $\entrywisemax($ \\
Algorithm & CPU Time & Wall-Clock & $\|A-U \Sigma V^*\|_2$ &
$\phantom{(}|U^*U-I|)$ & $\phantom{(}|V^*V-I|)$ \\
\noalign{\smallskip}\hline\noalign{\smallskip}
7 & 4.78E+02 & 7.41E+02 & 3.49E-15 & 2.44E-15 & 1.11E-15 \\
8 & 4.50E+02 & 7.43E+02 & 3.14E-15 & 2.11E-15 & 9.99E-16 \\
pre-existing & 7.99E+02 & 8.01E+02 & 1.09E-15 & 1.55E-15 & 5.55E-16 \\
\noalign{\smallskip}\hline
\end{tabular}
}
\end{table}

\begin{table}
\caption{$m =$ 10,000; $n =$ 2,000; $l = $ 20; $i = $ 2;
restricted to ten times fewer executors;
Appendix~\ref{devil} defines the singular values of the matrix being processed}
\label{tab56tiny18new}
{
\small
\begin{tabular}{rccccc}
\hline\noalign{\smallskip}
&&&& $\entrywisemax($ & $\entrywisemax($ \\
Algorithm & CPU Time & Wall-Clock & $\|A-U \Sigma V^*\|_2$ &
$\phantom{(}|U^*U-I|)$ & $\phantom{(}|V^*V-I|)$ \\
\noalign{\smallskip}\hline\noalign{\smallskip}
7 & 1.31E+02 & 1.26E+02 & 2.25E-15 & 9.78E-16 & 1.11E-15 \\
8 & 1.14E+02 & 1.26E+02 & 8.33E-15 & 1.78E-15 & 1.55E-15 \\
pre-existing & 1.66E+02 & 1.47E+02 & 7.80E-16 & 8.88E-16 & 8.88E-16 \\
\noalign{\smallskip}\hline
\end{tabular}
}
\end{table}

\begin{table}
\caption{Timings for $l = $ 10; $i = $ 2;
restricted to ten times fewer executors;
Appendix~\ref{devil} defines the singular values of the matrix being processed}
\label{partialSVDtime18new}
{
\small
\begin{tabular}{rrrcc}
\hline\noalign{\smallskip}
Algorithm & $m$ & $n$ & CPU Time & Wall-Clock \\
\noalign{\smallskip}\hline\noalign{\smallskip}
7 & 100,000 & 100,000 & 1.43E+04 & 1.01E+04 \\
8 & 100,000 & 100,000 & 1.41E+04 & 1.11E+04 \\
\noalign{\smallskip}\hline\noalign{\smallskip}
7 & 1,000,000 & 10,000 & 1.17E+04 & 1.45E+04 \\
8 & 1,000,000 & 10,000 & 1.13E+04 & 1.58E+04 \\
\noalign{\smallskip}\hline\noalign{\smallskip}
7 & 100,000 & 10,000 & 1.24E+03 & 1.11E+03 \\
8 & 100,000 & 10,000 & 1.16E+03 & 1.42E+03 \\
\noalign{\smallskip}\hline
\end{tabular}
}
\end{table}

\begin{table}
\caption{Errors for $l = $ 10; $i = $ 2;
restricted to ten times fewer executors;
Appendix~\ref{devil} defines the singular values of the matrix being processed}
\label{partialSVDerr18new}
{
\small
\begin{tabular}{rrrccc}
\hline\noalign{\smallskip}
&&&& $\entrywisemax($ & $\entrywisemax($ \\
Algorithm & $m$ & $n$ & $\|A-U \Sigma V^*\|_2$ &
$\phantom{(}|U^*U-I|)$ & $\phantom{(}|V^*V-I|)$ \\
\noalign{\smallskip}\hline\noalign{\smallskip}
7 & 100,000 & 100,000 & 3.26E-15 & 8.88E-16 & 1.33E-15 \\
8 & 100,000 & 100,000 & 3.14E-15 & 1.00E-15 & 1.01E-15 \\
\noalign{\smallskip}\hline\noalign{\smallskip}
7 & 1,000,000 & 10,000 & 2.45E-15 & 3.11E-15 & 5.77E-16 \\
8 & 1,000,000 & 10,000 & 4.20E-15 & 3.11E-15 & 9.99E-16 \\
\noalign{\smallskip}\hline\noalign{\smallskip}
7 & 100,000 & 10,000 & 1.72E-15 & 1.55E-15 & 1.11E-15 \\
8 & 100,000 & 10,000 & 2.10E-15 & 2.22E-15 & 8.88E-16 \\
\noalign{\smallskip}\hline
\end{tabular}
}
\end{table}

\clearpage

\section{Timings for generating the test matrices}
\label{gentimes}

For comparative purposes, Tables~\ref{generation}--\ref{generation3} list
the times required to generate~(\ref{originalmat}) with~(\ref{testS})
or~(\ref{testSl}) using the settings in Table~\ref{sparkset}.

\begin{table}
\caption{Timings for generating~(\ref{originalmat}) with~(\ref{testS})}
\label{generation}
{
\small
\begin{tabular}{rrcc}
\hline\noalign{\smallskip}
$m$ & $n$ & CPU Time & Wall-Clock \\
\noalign{\smallskip}\hline\noalign{\smallskip}
1,000,000 & 2,000 & 4.76E+03 & 3.91E+03 \\
  100,000 & 2,000 & 4.50E+02 & 2.48E+02 \\
   10,000 & 2,000 & 5.00E+01 & 2.60E+01 \\
\noalign{\smallskip}\hline
\end{tabular}
}
\end{table}

\begin{table}
\caption{Timings for generating~(\ref{originalmat}) with~(\ref{testSl})
and $l = 20$}
\label{generation2}
{
\small
\begin{tabular}{rrcc}
\hline\noalign{\smallskip}
$m$ & $n$ & CPU Time & Wall-Clock \\
\noalign{\smallskip}\hline\noalign{\smallskip}
1,000,000 & 2,000 & 5.61E+02 & 1.37E+03 \\
  100,000 & 2,000 & 6.30E+01 & 7.80E+01 \\
   10,000 & 2,000 & 8.00E+00 & 1.70E+01 \\
\noalign{\smallskip}\hline
\end{tabular}
}
\end{table}

\begin{table}
\caption{Timings for generating~(\ref{originalmat}) with~(\ref{testSl})
and $l = 10$}
\label{generation3}
{
\small
\begin{tabular}{rrcc}
\hline\noalign{\smallskip}
$m$ & $n$ & CPU Time & Wall-Clock \\
\noalign{\smallskip}\hline\noalign{\smallskip}
  100,000 & 100,000 & 7.30E+01 & 7.60E+01 \\
1,000,000 &  10,000 & 4.93E+02 & 1.79E+03 \\
  100,000 &  10,000 & 4.20E+01 & 5.20E+01 \\
\noalign{\smallskip}\hline
\end{tabular}
}
\end{table}

\clearpage

\bibliographystyle{spmpsci}      % mathematics and physical sciences
\bibliography{paper}   % name your BibTeX data base

\end{document}